\documentclass[12pt]{article}
\usepackage{lscape}
\usepackage{citesort}
\usepackage{amssymb}
\usepackage{appendix}
\usepackage{multirow}

\begin{document}

\def\qq{\langle \bar q q \rangle}
\def\uu{\langle \bar u u \rangle}
\def\dd{\langle \bar d d \rangle}
\def\sp{\langle \bar s s \rangle}
\def\GG{\langle g_s^2 G^2 \rangle}
\def\Tr{\mbox{Tr}}
\def\figt#1#2#3{
        \begin{figure}
        $\left. \right.$
        \vspace*{-2cm}
        \begin{center}
        \includegraphics[width=10cm]{#1}
        \end{center}
        \vspace*{-0.2cm}
        \caption{#3}
        \label{#2}
        \end{figure}
	}
	
\def\figb#1#2#3{
        \begin{figure}
        $\left. \right.$
        \vspace*{-1cm}
        \begin{center}
        \includegraphics[width=10cm]{#1}
        \end{center}
        \vspace*{-0.2cm}
        \caption{#3}
        \label{#2}
        \end{figure}
                }

\def\ds{\displaystyle}
\def\beq{\begin{equation}}
\def\eeq{\end{equation}}
\def\bea{\begin{eqnarray}}
\def\eea{\end{eqnarray}}
\def\beeq{\begin{eqnarray}}
\def\eeeq{\end{eqnarray}}
\def\ve{\vert}
\def\vel{\left|}
\def\ver{\right|}
\def\nnb{\nonumber}
\def\ga{\left(}
\def\dr{\right)}
\def\aga{\left\{}
\def\adr{\right\}}
\def\lla{\left<}
\def\rra{\right>}
\def\rar{\rightarrow}
\def\lrar{\leftrightarrow}  
\def\nnb{\nonumber}
\def\la{\langle}
\def\ra{\rangle}
\def\ba{\begin{array}}
\def\ea{\end{array}}
\def\tr{\mbox{Tr}}
\def\ssp{{\Sigma^{*+}}}
\def\sso{{\Sigma^{*0}}}
\def\ssm{{\Sigma^{*-}}}
\def\xis0{{\Xi^{*0}}}
\def\xism{{\Xi^{*-}}}
\def\qs{\la \bar s s \ra}
\def\qu{\la \bar u u \ra}
\def\qd{\la \bar d d \ra}
\def\qq{\la \bar q q \ra}
\def\gGgG{\la g^2 G^2 \ra}
\def\GG{\langle g_s^2 G^2 \rangle}
\def\g5{\gamma_5 \not\!q}
\def\x{\gamma_5 \not\!x}
\def\g5{\gamma_5}
\def\sb{S_Q^{cf}}
\def\sd{S_d^{be}}
\def\su{S_u^{ad}}
\def\sbp{{S}_Q^{'cf}}
\def\sdp{{S}_d^{'be}}
\def\sup{{S}_u^{'ad}}
\def\ssp{{S}_s^{'??}}

\def\sig{\sigma_{\mu \nu} \gamma_5 p^\mu q^\nu}
\def\fo{f_0(\frac{s_0}{M^2})}
\def\ffi{f_1(\frac{s_0}{M^2})}
\def\fii{f_2(\frac{s_0}{M^2})}
\def\O{{\cal O}}
\def\sl{{\Sigma^0 \Lambda}}
\def\es{\!\!\! &=& \!\!\!}
\def\ap{\!\!\! &\approx& \!\!\!}
\def\ar{&+& \!\!\!}
\def\arrr{\!\!\!\! &+& \!\!\!}
\def\ek{&-& \!\!\!}
\def\vev{&\vert& \!\!\!}
\def\kek{\!\!\!\!&-& \!\!\!}
\def\cp{&\times& \!\!\!}
\def\se{\!\!\! &\simeq& \!\!\!}
\def\eqv{&\equiv& \!\!\!}
\def\kpm{&\pm& \!\!\!}
\def\kmp{&\mp& \!\!\!}
\def\mcdot{\!\cdot\!}
\def\erar{&\rightarrow&}


\def\simlt{\stackrel{<}{{}_\sim}}
\def\simgt{\stackrel{>}{{}_\sim}}


\title{
         {\Large
                 {\bf
Strong coupling constant of negative parity octet baryons with
light pseudoscalar mesons in light cone QCD sum rules
                 }
         }
      }

\author{\vspace{1cm}\\
{\small T. M. Aliev$^1$ \thanks {e-mail: taliev@metu.edu.tr}~\footnote{permanent address:Institute of
Physics,Baku,Azerbaijan}\,\,,
M. Savc{\i}$^1$ \thanks
{e-mail: savci@metu.edu.tr}} \\
{\small $^1$ Physics Department, Middle East Technical University,
06531 Ankara, Turkey }\\
}

\date{}

\begin{titlepage}
\maketitle
\thispagestyle{empty}

\begin{abstract}

The strong coupling constants of the $\pi$ and $K$ mesons with negative
parity octet baryons are estimated within the light cone QCD sum rules.
It is observed that all strong coupling constants, similar to the case for
the positive parity baryons,
can be described in terms of three invariant functions, where two
of them correspond to the well known $F$ and $D$ couplings in the $SU(3)_f$
symmetry, and the third function describes the  $SU(3)_f$ symmetry
violating effects. We compare our predictions on the strong coupling
constants of pseudoscalar mesons of negative parity baryons with those
corresponding to the strong coupling constants for the positive parity baryons.

\end{abstract}

~~~PACS numbers: 14.20.Dh, 14.40.Be, 13.75.Gx, 11.55.Hx

\end{titlepage}

\section{Introduction}

The hadron--meson strong coupling constants play central role in the analysis of
the existing experimental results of hadron physics. Among the strong
coupling constants of baryons with mesons only nucleon--pion coupling
constant is well determined from the experiments available at present.
However, the situation for the $K$ meson is not that simple. In reproducing the
experimental results for the kaon--nucleon scattering and kaon photoproduction
many phenomenologically unknown coupling constants are necessary
(see \cite{Rvsf01}). Moreover, as far as
negative parity baryons are concerned there exists limited experimental
data. Understanding dynamics of the negative parity baryons requires
formidable efforts, both from the theoretical and the experimental sides.

During the recent years considerable progress has been made in determination of
the masses, magnetic moments of the negative parity baryons (see for
example,
\cite{Rvsf02,Rvsf03,Rvsf04,Rvsf05,Rvsf06,Rvsf07,Rvsf08,Rvsf09,Rvsf10,Rvsf11}).
Following these works, light cone QCD sum rules is also applied in studying
the electromagnetic transition form factors for the $\gamma^\ast N \to
N(1520)$ \cite{Rvsf12}. The next step in investigation of the
properties of the negative parity baryons would be the analysis of their coupling
constants with light pseudoscalar mesons. These coupling constants are the
key parameters for studying the physics of negative parity baryons.

The present work is devoted to the study of the strong coupling constants of
negative parity octet baryons $(O^\ast)$ with light pseudoscalar mesons
within the light cone QCD sum rules. The
work is arranged as follows. In Section 2 the relevant sum rules for the
strong coupling constants of the negative parity baryons with the light
pseudoscalar mesons are derived. In Section 3 the numerical analysis of the
sum rules obtained for the strong coupling constant
$g_{O^\ast O^\ast {\cal P}}$ is performed.

\section{Light cone sum rules for the $g_{O^\ast O^\ast {\cal P}}$ coupling
constant}

In order to determine $g_{O^\ast O^\ast {\cal P}}$ coupling constant
we start by considering the following correlation function,
\bea
\label{evsf01}
\Pi(p,q) = i \int d^4x e^{ipx} \lla {\cal P}(q) \vel \mbox{T} 
\Big\{ \eta_{B_2} (x) \bar{\eta}_{B_1} (0)\ver 0 \rra\,,
\eea
where $B_2(B_1)$ are the final (initial) baryons, $\eta$ is the
interpolating current of the corresponding baryon, $q$ is the momentum of
the pseudoscalar meson. Following the strategy of the QCD sum rules, we
must first compute the correlation function (\ref{evsf01}) in two different
kinematical domains. If the initial and final baryons are close to the mass
shell of the corresponding baryons, i.e., $p^2 \simeq m_{B_2}^2$ and $(p+q)^2
\simeq m_{B_1}^2$ the hadronic part of the correlation function will be
dominated by the processes $O^\ast \to O^\ast {\cal P}$ and $O^\ast \to
O {\cal P}$, i.e., octet baryons of both parities contribute to the
correlation function.

It should be noted here that in the case of nucleons, in addition to the
$N(938)$ state the next positive parity $N(1440)$ state also gives
contribution to the correlation function, while such states are absent in
all other members of octet baryons. This fact makes the calculation of the
nucleon coupling constant $g_{N^\ast N^\ast\pi}$ more challenging, whose
technical details is discussed in \cite{Rvsf13}.   

The general form of the octet baryon currents are,
\bea
\label{evsf02}
\eta_{\Sigma^+} \es 2 \varepsilon^{abc} \sum_{\ell=1}^2 \left(u^{aT} C A_1^\ell
s^b\right) A_2^\ell u^c\,,\nnb \\
\eta_{\Sigma^+} \es \eta_{\Sigma^+} (u \to d)\,,\nnb \\
\eta_{\Sigma^0} \es \sqrt{2} \varepsilon^{abc} \sum_{\ell=1}^2 \left\{\left(u^{aT} C
A_1^\ell s^b\right) A_2^\ell d^c + \left(d^{aT} C A_1^\ell s^b\right)
A_2^\ell u^c\right\}\,,\nnb \\
\eta_{\Xi^0} \es \eta_{\Sigma^+} (u \to s)\,,\nnb \\
\eta_{\Xi^-} \es \eta_{\Sigma^-} (d \to s)\,.
\eea
Here $C$ is the charge conjugation operator, $A_1^1=I$,
$A_1^2=A_2^1=\gamma_5$, and $A_2^2=\beta$ where $\beta$ is an arbitrary
auxiliary parameter. It is shown in \cite{Rvsf14} that the current of
$\Lambda$ baryon can be obtained from the $\Sigma^0$ current with the help of
following relations,
\bea
\label{evsf03}
2 \eta_{\Sigma^0} (d \lrar s) + \eta_{\Sigma^0} \es -\sqrt{3}
\eta_\Lambda~~\mbox{or,}~ \nnb \\
2 \eta_{\Sigma^0} (u \lrar s) - \eta_{\Sigma^0} \es -\sqrt{3} \eta_\Lambda\,.
\eea
Note that these currents interact with both positive and negative parity
baryons.

In order to find the expression of the correlation function from the
hadronic part the correlation function (\ref{evsf01}) needs to be saturated with
the hadronic states as follows,
\bea
\label{evsf04}
\Pi(p,q)  = \sum_{i,j} {\lla 0 \vel \eta_2 \ver B_2^i(p) \rra \lla B_2^i(p)
{\cal P}(q) \ve B_1^j (p+q)
\rra \lla B_1^j(p+q) \vel \bar{\eta}_1 \ver 0 \rra \over
(p_i^2-m_{2i}^2)[(p+q)^2 - m_{1j}^2] }\,,
\eea
where the summation over positive and negative parity baryons are implemented by
the subindices $i$ and $j$. The matrix elements entering into  Eq. (\ref{evsf01})
are determined as,
\bea
\label{evsf05}
\lla 0 \vel \eta \ver B^+ (p,s)\rra \es \lambda_+
u_+(p,s)\,, \nnb \\
\lla 0 \vel \eta \ver B^- (p,s)\rra \es \lambda_-
\gamma_5 u_+(p,s)\,, \nnb \\
\lla B_2(p)^+ {\cal P}\ve B_1^+(p+q)  \rra \es i g_{++} \bar{u}_+ \gamma_5 u_+\,, \nnb\\
\lla B_2(p)^- {\cal P}\ve B_1^-(p+q)  \rra \es i g_{--} \bar{u}_- \gamma_5 u_-\,, \nnb\\
\lla B_2(p)^+ {\cal P}\ve B_1^-(p+q)  \rra \es i g_{-+} \bar{u}_+ u_-\,, \nnb\\
\lla B_2(p)^- {\cal P}\ve B_1^+(p+q)  \rra \es i g_{+-} \bar{u}_- u_+\,,
\eea

Using these matrix elements and performing the summation over the spins of
Dirac spinors, the hadronic part of the correlation function can be written
as,
\bea
\label{evsf06}
\Pi \es
i g_{++} \lambda_{1+} \lambda_{2+} {(\not\!{p} + m_{2+}) \gamma_5 (\not\!{p} + \not\!{q} +
m_{1+}) \over [(p+q)^2 - m_{1+}^2] (p^2 - m_{2+}^2)} \nnb \\
\ek i g_{--} \lambda_{1-} \lambda_{2-} {\gamma_5 (\not\!{p} + m_{2-}) \gamma_5 (\not\!{p} + \not\!{q} +
m_{1-}) \gamma_5  \over [(p+q)^2 - m_{1-}^2] (p^2 - m_{2-}^2)} \nnb \\
\ek i g_{-+} \lambda_{1-} \lambda_{2+} {(\not\!{p} + m_{2+}) (\not\!{p} + \not\!{q} +
m_{1-}) \gamma_5 \over [(p+q)^2 - m_{1-}^2] (p^2 - m_{2+}^2)} \nnb \\
\ar i g_{+-} \lambda_{1+} \lambda_{2-} {\gamma_5 (\not\!{p} + m_{2-}) (\not\!{p} + \not\!{q} +          
m_{1+}) \over [(p+q)^2 - m_{1+}^2] (p^2 - m_{2-}^2)}\,.  
\eea

Few words are in order about the contributions that are expected to come from single
pole terms for each state. In principle the strong coupling
constant $g_{{\cal P}O^\ast O^\ast}$ has $q^2$ momentum dependence since
it contains pseudoscalar meson form factor, which is described by the
single pole contributions. In the framework
of the light cone QCD sum rules approach we have used,
the pseudoscalar mesons are assumed on the mass shell, i.e., $q^2 =
m_{\cal P}^2$, and therefore the strong coupling constant 
$g_{{\cal P}O^\ast O^\ast}$ has no $q^2$ dependency.
Under this condition the correlation function depends on
two variables, namely  $p^2$ and $(p+q)^2$, and have double poles with respect
to these variables. In general single pole terms appear in subtraction
procedure, which makes double dispersion integral finite.

In determination of the strong coupling constants of pseudoscalar mesons
with negative parity octet baryons, the double Borel transformation is performed
with respect to the variables $-p^2$ and $-(p+q)^2$, and
the single pole terms that depend on only either one of these variables vanish.

In order to calculate the light pseudoscalar meson--negative parity hyperon
coupling constant $g_{--}$, the coupling constant $g_{++}$  between the
positive--positive
parity baryons, as well as the coupling constants $g_{-+}$ and $g_{+-}$
between the negative--positive
and positive--negative parity hyperons, respectively, must be eliminated from the
four coupled linear
equations. For this purpose we need to calculate the correlation function
from the QCD side.

Before proceeding further, 
following the approach given in \cite{Rvsf15} and \cite{Rvsf16}, we first
derive the corresponding relations among the correlation functions
describing the coupling constants of the negative parity octet baryons with
light pseudoscalar mesons. It is shown in these studies that all correlation
functions can be expressed in terms of three independent functions, where
two of them correspond to the $SU(3)_f$ symmetry limit and the third function
takes $SU(3)_f$ violation into account. This is quite an exciting result
since the coupling constants of all pseudoscalar mesons with baryons are all
expressed in terms of $F$ and $D$ constants in the $SU(3)_f$ symmetry.
Another advantage of this approach is that the relations among the invariant
functions are structure independent. It can be shown by a specific example 
how these three invariant functions appear in the approach we consider.
Consider the invariant function
\bea
\label{evsf07}
\Pi = g_{\pi^0\bar{u}u} \Pi_1 (u,d,s) + g_{\pi^0\bar{d}d} \Pi_1^\prime (u,d,s) +
g_{\pi^0\bar{s}s} \Pi_2 (u,d,s)\,,
\eea
for the $\Sigma^0 \to \Sigma^0 \pi^0$ transition,
where $\Sigma^0$ is the common notation for the positive and negative parity
baryons. The quark content of the $\pi^0$ meson is formally written as,
\bea
J = \sum_{q=u,d,s} g_{\pi^0\bar{q}q} \bar{q} \gamma_5 q\,,\nnb
\eea
and for the $\pi^0$ meson $g_{\pi^0\bar{u}u}=-g_{\pi^0\bar{d}d} =
1/\sqrt{2}$ and $g_{\pi^0\bar{s}s}=0$. Since the interpolating current of
$\Sigma^0$ is symmetric under the exchange $u \lrar d$, i.e.,
$\Pi_1^\prime (u,d,s) = \Pi_1(d,u,s)$, so can write,
\bea
\label{evsf08}
\Pi^{\Sigma^0\to \Sigma^0 \pi^0} = {1\over \sqrt{2}} \Big[\Pi_1(u,d,s) -
\Pi_1(d,u,s)\Big]\,,
\eea
and obviously this invariant function is equal to zero at $SU(2)_f$ limit.
For the convenience we introduce the following formal notation,
\bea
\label{evsf09}
\Pi_1(u,d,s) \es \lla \bar{u}u \vel \Sigma^0 \bar{\Sigma}^0 \ver 0 \rra\,, \nnb\\
\Pi_2(u,d,s) \es \lla \bar{s}s \vel \Sigma^0 \bar{\Sigma}^0 \ver 0 \rra\,. \nnb\\ 
\eea
Making the replacement $d \to u$ in $\Pi_1$ and using $\Sigma^0(d\to u) =
- \sqrt{2} \Sigma^+$, one can easily show that,
\bea
\lla \bar{u}u \vel \Sigma^+ \bar{\Sigma}^+ \ver 0 \rra = 2 \Pi_1 (u,u,s)\,,\nnb
\eea
so that we can write
\bea
\label{evsf10}
\Pi^{\Sigma^+ \to \Sigma^+ \pi^0} = \sqrt{2} \Pi_1(u,u,s)\,.
\eea

The invariant functions involving $\Lambda$ baryon appear with the ones
involving $\Sigma^0$ baryon (see Eq. (\ref{evsf03})) and therefore it is
impossible to write them using only $\Pi_1$ and $\Pi_2$. Hence we need to
introduce one
more independent function in order to separate the contributions coming from
$\Lambda$ and $\Sigma^0$ which can be written as,
\bea
\Pi_3(u,d,s) = \Pi^{\Sigma^0 \to \Xi^- K^+} = - \lla u \bar{s} \vel \Xi^-
\Sigma^0 \ver 0 \rra\,.\nnb
\eea
The relations among invariant functions in the isospin symmetry limit
are given in the Appendix (see also \cite{Rvsf15}).

We now proceed to calculate the correlation functions from the QCD side. As
already mentioned, all considered correlation functions can be written
in terms of three invariant functions, which can be determined from the
$\Sigma^0 \to \Sigma^0 \pi^0$ and $\Sigma^0 \to \Xi^- K^+$ transitions.
Therefore in order to find the relations among correlation functions it is
enough to calculate  correlation functions responsible for these decays.
Theoretical part of the correlation function describing the above transitions
can be calculated at the deep Eucledian region $p^2 \ll 0$ and $(p+q)^2 \ll 0$
using the OPE over twists. In these calculations there appear matrix elements
of nonlocal operators between vacuum and pseudoscalar meson states, such as,
$\lla {\cal P} (q) \vel \bar{q}_1 (x_1) \Gamma q_2(x_2) \ver 0 \rra$ or
$\lla {\cal P} (q) \vel \bar{q}_1 (x_1) \Gamma G q_2(x_2) \ver 0 \rra$,
where $\Gamma$ is the gluon field strength tensor. It should be noted here
that contributions coming from four--particle nonlocal operators  like
$\bar{q} GG q$, $\bar{q}q\bar{q}q$ are all neglected. Formally, neglecting
the contributions of these operators can be justified on the basis of an
expansion in the conformal spin method \cite{Rvsf17}. Therefore we restrict ourselves
to the contributions of the two-- and three--particle pseudoscalar meson
DAs up to twist four. In these calculations we also need the light quark
propagator in the presence of background gluon field whose expression is
given in \cite{Rvsf18},
\bea
S_q(x) \es {i \rlap/x\over 2\pi^2 x^4} [x,0]
- {i g_s \over 16 \pi^2 x^2} \int_0^1 du [x,ux] \bigg[\bar{u} \rlap/{x}
\sigma_{\alpha \beta} +  u \sigma_{\alpha \beta} \bigg]
G^{\alpha \beta} [ux,0] + \cdots\nnb 
\eea
where the aberration $[x,y]$ means,
\bea
[x,y] = P exp\Big[ i \int dt (x-y)_\mu g B^\mu (tx-\bar{t}y) \Big]\,,\nnb
\eea
for the path ordered exponent, and $B$ is the gluon field. The calculations
are performed in the Fock--Schwinger gauge, i.e., $x_\mu B^\mu = 0$. As the
result the massless quark propagator can be written as,
\bea
S_q(x) \es {i \rlap/x\over 2\pi^2 x^4} -
ig_s \int_0^1 du \Bigg[ {\bar{u}\rlap/x  \sigma_{\alpha \beta}
G^{\alpha \beta} (ux) \over 16 \pi^2 x^2} -
{u x_\alpha \gamma_\beta G^{\alpha \beta} (ux) \over 4\pi^2 x^2}\Bigg]\,.\nnb
\eea
If the mass of the light quark is taken into account this propagator is
modified as,
\bea
S_q(x) \es {i \rlap/x\over 2\pi^2 x^4} - {m_q\over 4 \pi^2 x^2}
- i g_s  \int_0^1 du \Bigg[{\bar{u} \rlap/{x}
\sigma_{\alpha \beta} G^{\alpha \beta} (ux) \over 16 \pi^2 x^2}
- {u x_\alpha \gamma_\beta G^{\alpha \beta} (ux)\over 4 \pi^2 x^2} \nnb \\
\ek {i m_q x^2 \over 32} \sigma_{\alpha \beta} G^{\alpha \beta} (ux)
\left( \ln {-x^2 \Lambda^2\over 4}  +
 2 \gamma_E \right) \Bigg]\,, \nnb
\eea
where $\gamma_E\approx 0.577$ is the Euler constant, and $\Lambda$ is the
scale that separates the short and long distance domains. We choose this
parameter as the factorization scale that has the value $\Lambda =1\,GeV$
(see for example \cite{Rvsf19}).

The matrix elements of the nonlocal operators between one pseudoscalar and
vacuum states are defined in terms of the meson distribution amplitudes as,
\cite{Rvsf20,Rvsf21,Rvsf22}
\bea
\label{evsf11} 
\lla {\cal P}(q)\vel \bar q_1(x) \gamma_\mu \gamma_5 q_1(0)\ver 0 \rra \es
-i f_{\cal P} q_\mu  \int_0^1 du  e^{i \bar u q x}
    \left( \varphi_{\cal P}(u) + {1\over 16} m_{\cal P}^2
x^2 {\Bbb{A}}(u) \right) \nnb \\
\ek {i\over 2} f_{\cal P} m_{\cal P}^2 {x_\mu\over qx}
\int_0^1 du e^{i \bar u qx} {\Bbb{B}}(u)\,,\nnb \\
\lla {\cal P}(q)\vel \bar q_1(x) i \gamma_5 q_2(0)\ver 0 \rra \es
\mu_{\cal P} \int_0^1 du e^{i \bar u qx} \phi_P(u)\,,\nnb \\
\lla {\cal P}(q)\vel \bar q_1(x) \sigma_{\alpha \beta} \gamma_5 q_2(0)\ver 0 \rra \es
{i\over 6} \mu_{\cal P} \left( 1 - \widetilde{\mu}_{\cal P}^2 \right)
\left( q_\alpha x_\beta - q_\beta x_\alpha\right)
\int_0^1 du e^{i \bar u qx} \phi_\sigma(u)\,,\nnb \\
\lla {\cal P}(q)\vel \bar q_1(x) \sigma_{\mu \nu} \gamma_5 g_s
G_{\alpha \beta}(v x) q_2(0)\ver 0 \rra \es i \mu_{\cal P} \left[
q_\alpha q_\mu \left( g_{\nu \beta} - {1\over qx}(q_\nu x_\beta +
q_\beta x_\nu) \right) \right. \nnb \\
\ek q_\alpha q_\nu \left( g_{\mu \beta} -
{1\over qx}(q_\mu x_\beta + q_\beta x_\mu) \right) \nnb \\
\ek q_\beta q_\mu \left( g_{\nu \alpha} - {1\over qx}
(q_\nu x_\alpha + q_\alpha x_\nu) \right) \nnb \\
\ar q_\beta q_\nu \left. \left( g_{\mu \alpha} -
{1\over qx}(q_\mu x_\alpha + q_\alpha x_\mu) \right) \right] \nnb \\
\cp \int {\cal D} \alpha e^{i (\alpha_{\bar q} +
v \alpha_g) qx} {\cal T}(\alpha_i)\,,\nnb \\
\lla {\cal P}(q)\vel \bar q_1(x) \gamma_\mu \gamma_5 g_s
G_{\alpha \beta} (v x) q_2(0)\ver 0 \rra \es q_\mu (q_\alpha x_\beta -
q_\beta x_\alpha) {1\over qx} f_{\cal P} m_{\cal P}^2
\int {\cal D}\alpha e^{i (\alpha_{\bar q} + v \alpha_g) qx}
{\cal A}_\parallel (\alpha_i) \nnb \\
\ar \left[q_\beta \left( g_{\mu \alpha} - {1\over qx}
(q_\mu x_\alpha + q_\alpha x_\mu) \right) \right. \nnb \\
\ek q_\alpha \left. \left(g_{\mu \beta}  - {1\over qx}
(q_\mu x_\beta + q_\beta x_\mu) \right) \right]
f_{\cal P} m_{\cal P}^2 \nnb \\
\cp \int {\cal D}\alpha e^{i (\alpha_{\bar q} + v \alpha _g)
q x} {\cal A}_\perp(\alpha_i)\,,\nnb \\
\lla {\cal P}(q)\vel \bar q_1(x) \gamma_\mu i g_s G_{\alpha \beta}
(v x) q_2(0)\ver 0 \rra \es q_\mu (q_\alpha x_\beta - q_\beta x_\alpha)
{1\over qx} f_{\cal P} m_{\cal P}^2 \int {\cal D}\alpha e^{i (\alpha_{\bar q} +
v \alpha_g) qx} {\cal V}_\parallel (\alpha_i) \nnb \\
\ar \left[q_\beta \left( g_{\mu \alpha} - {1\over qx}
(q_\mu x_\alpha + q_\alpha x_\mu) \right) \right. \nnb \\
\ek q_\alpha \left. \left(g_{\mu \beta}  - {1\over qx}
(q_\mu x_\beta + q_\beta x_\mu) \right) \right] f_{\cal P} m_{\cal P}^2 \nnb \\
\cp \int {\cal D}\alpha e^{i (\alpha_{\bar q} +
v \alpha _g) q x} {\cal V}_\perp(\alpha_i)\,,
\eea
where
\bea
\mu_{\cal P} = f_{\cal P} {m_{\cal P}^2\over m_{q_1} + m_{q_2}}\,,~~~~~
\widetilde{\mu}_{\cal P} = {m_{q_1} + m_{q_2} \over m_{\cal P}}\,, \nnb
\eea
and ${\cal D}\alpha = d\alpha_{\bar q} d\alpha_q d\alpha_g
\delta(1-\alpha_{\bar q} - \alpha_q - \alpha_g)$ is the measure. Here 
$\varphi_{\cal P}(u)$ is the leading twist--two, $\phi_P(u)$, $\phi_\sigma(u)$,
${\cal T}(\alpha_i)$ are the twist--three, and
$\Bbb{A}(u)$, $\Bbb{B}(u)$, ${\cal A}_\perp(\alpha_i),$ ${\cal A}_\parallel(\alpha_i),$
${\cal V}_\perp(\alpha_i)$ and ${\cal V}_\parallel(\alpha_i)$
are the twist--four DAs, respectively. The explicit forms of these matrix
elements are given in the following section. 

In order to proceed with the calculation of the strong coupling constant
between negative parity baryons with light pseudoscalar mesons, we need to
solve Eq. (\ref{evsf05}) for $g_{--}$. We see from this equation that four
linearly independent equations are needed for determination of the strong
coupling constant $g_{--}$, and hence we choose four different
Lorentz structures, $\rlap/{p}\rlap/{q}\gamma_5$,
$\rlap/{p}\gamma_5$, $\rlap/{q}\gamma_5$ and $\gamma_5$ which correspond to
the invariant functions $\Pi_i^{(1)}$, $\Pi_i^{(2)}$, $\Pi_i^{(3)}$, and
$\Pi_i^{(4)}$ respectively. Here the subindex $i$
runs over the invariant functions $\Pi_1$, $\Pi_2$, $\Pi_3$ as are
introduced earlier. 

In order to find sum rules for the strong coupling constant of the negative
parity octet baryons with pseudoscalar mesons, we should match both
representations of the correlation function from the hadronic and QCD sides.
Performing the double Borel transformation over the variables $-p^2$ and
$-(p+q)^2$ the higher states and continuum contributions are suppressed. The
contributions are calculated by using the quark--hadron duality, i.e., above
some threshold, the hadronic spectral density is equal to the spectral
density calculated in terms of quark--gluon degrees of freedom.
Following the Borel transformation, the continuum subtraction procedure is applied
whose details are given in \cite{Rvsf21}. Setting $M_1^2=M_2^2=2  M^2$ and
$u_0=1/2$ the continuum subtraction procedure can be done using the
relation,
\bea
(M^2)^n = {1\over \Gamma(n)} \int_0^{s_0} ds\,e^{-s/M^2} s^{n-1}\,.\nnb
\eea
The continuum subtraction is not performed for the higher twist terms due to the
fact that their contributions are known to be small (for more details see
\cite{Rvsf21}).

Having implemented the Borel transformation and continuum subtraction procedures we obtain four
equations, which correspond to the transitions between positive--positive,
negative--negative, positive--negative and negative--positive parity
baryons.
\bea
&&-A+B-C+D = \Pi_i^{B(1)}\,,\nnb\\
&&-m_{2+} A - m_{2-}B -m_{2+}C-m_{2-}D = \Pi_i^{B(2)}\,,\nnb\\
&&(m_{1+}-m_{2-}) A + (m_{1-}-m_{2-}) B - (m_{1-}+m_{2+}) C -
(m_{1+}+m_{2-}) D = \Pi_i^{B(3)}\,,\nnb\\
&&m_{2+} (m_{1+} - m_{2+}) A - m_{2-} (m_{1-} - m_{2-}) B -
m_{2+} ( m_{1-} + m_{2+}) C \nnb \\
&& + m_{2-} (m_{1+} + m_{2-}) D = \Pi_i^{B(4)}\,,
\eea
where
\bea
\label{evsf12}
A \es g_{++} \lambda_{1+} \lambda_{2+} e^{-(m_{1+}^2+m_{2+}^2)/2 M^2}\,,\nnb\\
B \es g_{--} \lambda_{1-} \lambda_{2-} e^{-(m_{1-}^2+m_{2-}^2)/2 M^2}\,,\nnb\\
C \es g_{-+} \lambda_{1-} \lambda_{2+} e^{-(m_{1-}^2+m_{2+}^2)/2 M^2}\,,\nnb\\
D \es g_{+-} \lambda_{1+} \lambda_{2-} e^{-(m_{1+}^2+m_{2-}^2)/2 M^2}\,.\nnb
\eea
In order obtain the $g_{{\cal P}O^\ast O^\ast}$ coupling constant we
need to determine the function $B$ only.
Solving these four equations the strong coupling constant is found
to be,
\bea
\label{evsf13}
g_{--} \es {e^{(m_{1-}^2+m_{2-}^2)/2M^2+m_{\cal P}^2/4M^2} \over
\lambda_{1-} \lambda_{2-} (m_{1-} + m_{1+}) (m_{2-} + m_{2+})} 
\Bigg\{ (m_{1+} + m_{2-}) m_2^+ \Pi_i^{B(1)} + 
m_{2+} \Pi_i^{B(2)} \nnb \\
\ek (m_{1+} + m_{2-}) \Pi_i^{B(3)} - 
\Pi_i^{B(4)}\Bigg\}\,,
\eea
where $\Pi_i^{B(j)}$ correspond to the invariant functions
$\Pi_i^{(j)}$ after the Borel transformation with respect to the variables
$-p^2$ and $-(p+q)^2$.
The explicit expressions of the invariant functions $\Pi^{B(i)}$
$(i=1,2,3,4)$ for any considered strong coupling constant between negative
parity baryons with light pseudoscalar mesons are quite lengthy, and hence
we do not present their explicit forms.

It follows from Eq. (\ref{evsf13}) that to able to estimate of the strong coupling constants
of negative parity octet baryons with light pseudoscalar mesons, residues of
negative parity octet baryons are needed. These residues are calculated in
\cite{Rvsf11,Rvsf12}  
       
\section{Numerical analysis}

This section is devoted to the numerical analysis of the sum rules
obtained in Section 3 for the strong coupling
constants of the light pseudoscalar mesons with negative parity octet
baryons. 
The sum rules of the aforementioned 
coupling constants contain the DAs of the pseudoscalar mesons as the input
parameters, and their explicit
forms are given below \cite{Rvsf20,Rvsf21,Rvsf22}. 

\bea
\label{evsf14}
\varphi_{\cal P}(u) \es 6 u \bar u \left[ 1 + a_1^{\cal P} C_1(2 u -1) +
a_2^{\cal P} C_2^{3/2}(2 u - 1) \right]\,,  \nnb \\
{\cal T}(\alpha_i) \es 360 \eta_3 \alpha_{\bar q} \alpha_q
\alpha_g^2 \left[ 1 + w_3 {1\over 2} (7 \alpha_g-3) \right]\,, \nnb \\
\phi_P(u) \es 1 + \left[ 30 \eta_3 - {5\over 2}
{1\over \mu_{\cal P}^2}\right] C_2^{1/2}(2 u - 1)\,,  \nnb \\
\ar \left( -3 \eta_3 w_3  - {27\over 20} {1\over \mu_{\cal P}^2} -
{81\over 10} {1\over \mu_{\cal P}^2} a_2^{\cal P} \right)
C_4^{1/2}(2u-1)\,, \nnb \\
\phi_\sigma(u) \es 6 u \bar u \left[ 1 + \left(5 \eta_3 - {1\over 2} \eta_3 w_3 -
{7\over 20}  \mu_{\cal P}^2 - {3\over 5} \mu_{\cal P}^2 a_2^{\cal P} \right)
C_2^{3/2}(2u-1) \right] \,, \nnb \\
{\cal V}_\parallel(\alpha_i) \es 120 \alpha_q \alpha_{\bar q} \alpha_g
\left( v_{00} + v_{10} (3 \alpha_g -1) \right) \,, \nnb \\
{\cal A}_\parallel(\alpha_i) \es 120 \alpha_q \alpha_{\bar q} \alpha_g
\left( 0 + a_{10} (\alpha_q - \alpha_{\bar q}) \right) \,, \nnb \\
{\cal V}_\perp (\alpha_i) \es - 30 \alpha_g^2\left[ h_{00}(1-\alpha_g) +
h_{01} (\alpha_g(1-\alpha_g)- 6 \alpha_q \alpha_{\bar q}) +
h_{10}(\alpha_g(1-\alpha_g) - {3\over 2} (\alpha_{\bar q}^2+
\alpha_q^2)) \right] \,, \nnb \\
{\cal A}_\perp (\alpha_i) \es 30 \alpha_g^2(\alpha_{\bar q} - \alpha_q)
\left[ h_{00} + h_{01} \alpha_g + {1\over 2} h_{10}(5 \alpha_g-3) \right] \,, \nnb \\
B(u)\es g_{\cal P}(u) - \varphi_{\cal P}(u) \,, \nnb \\
g_{\cal P}(u) \es g_0 C_0^{1/2}(2 u - 1) + g_2 C_2^{1/2}(2 u - 1) +
g_4 C_4^{1/2}(2 u - 1) \,, \nnb \\
\Bbb{A}(u) \es 6 u \bar u \left[{16\over 15} + {24\over 35} a_2^{\cal P}+
20 \eta_3 + {20\over 9} \eta_4 +
\left( - {1\over 15}+ {1\over 16}- {7\over 27}\eta_3 w_3 -
{10\over 27} \eta_4 \right) C_2^{3/2}(2 u - 1)  \right. \nnb \\
    \ar \left. \left( - {11\over 210}a_2^{\cal P} - {4\over 135}
\eta_3w_3 \right)C_4^{3/2}(2 u - 1)\right] \,, \nnb \\
\ar \left( -{18\over 5} a_2^{\cal P} + 21 \eta_4 w_4 \right)
\left[ 2 u^3 (10 - 15 u + 6 u^2) \ln u  \right. \nnb \\
\ar \left. 2 \bar u^3 (10 - 15 \bar u + 6 \bar u ^2) \ln\bar u +
u \bar u (2 + 13 u \bar u) \right]\,.
\eea
Here $C_n^k(x)$ are the Gegenbauer polynomials, and
\bea
\label{evsf15}
h_{00}\es v_{00} = - {1\over 3}\eta_4 \,, \nnb \\
a_{10} \es {21\over 8} \eta_4 w_4 - {9\over 20} a_2^{\cal P} \,, \nnb \\
v_{10} \es {21\over 8} \eta_4 w_4 \,, \nnb \\
h_{01} \es {7\over 4}  \eta_4 w_4  - {3\over 20} a_2^{\cal P} \,, \nnb \\
h_{10} \es {7\over 4} \eta_4 w_4 + {3\over 20} a_2^{\cal P} \,, \nnb \\
g_0 \es 1 \,, \nnb \\
g_2 \es 1 + {18\over 7} a_2^{\cal P} + 60 \eta_3  + {20\over 3} \eta_4 \,, \nnb \\
g_4 \es  - {9\over 28} a_2^{\cal P} - 6 \eta_3 w_3\,.
\eea
The values of the parameters $a_1^{\cal P}$, $a_2^{\cal P}$,
$\eta_3$, $\eta_4$, $w_3$, and $w_4$ entering Eq. (\ref{evsf14}) are listed in
Table (\ref{param}) for the pseudoscalar $\pi$, $K$ and $\eta$ mesons.

\begin{table}[h]
\def\bos{\lower 0.25cm\hbox{{\vrule width 0pt height 0.7cm}}}
\begin{center}
\begin{tabular}{|c|c|c|}
\hline\hline
        & \bos  $\pi$   &   $K$ \\
\hline
$a_1^{\cal P}$  & \bos   0 &   0.050 \\
\hline
$a_2^{\cal P}~\mbox{(set-1)}$  & \bos   0.11  &   0.15 \\
\hline
$a_2^{\cal P}~\mbox{(set-2)}$  &  \bos  0.25  &   0.27 \\
\hline
$\eta_3$    & \bos  0.015 &   0.015 \\
\hline
$\eta_4$    & \bos  10    &   0.6 \\
\hline
$w_3$       & \bos  $-3$    &   $-3$ \\
\hline
$w_4$       & \bos  0.2   &   0.2 \\
\hline \hline
\end{tabular}
\end{center}
\caption{Parameters of the wave function calculated at the renormalization scale $\mu = 1\,GeV$}
\label{param}
\end{table}

In addition to the DAs, these coupling constants contain also additional
parameters such as  quark condensates and magnetic susceptibility of quarks.
In the present analysis we use $\uu\ve_{\mu=1\,GeV} = \dd\ve_{\mu=1\,GeV} =
-(0.243)^3\,GeV^3$ \cite{Rvsf23}, $\sp\ve_{\mu=1\,GeV} = 0.8
\uu\ve_{\mu=1\,GeV}$, $m_0^2=(0.8\pm0.2)\,GeV^2$ \cite{Rvsf24}. Magnetic
susceptibility of the quarks are determined in the framework of the QCD sum rules
in \cite{Rvsf25,Rvsf26,Rvsf27} and in our work we use
$\chi(1\,GeV)=-2.85\,GeV^2$ predicted in \cite{Rvsf27}.

Besides these input parameters the sum rules contain three auxiliary parameters:
Borel mass square $M^2$, continuum threshold $s_0$ and the arbitrary
parameter $\beta$ in expressions of the interpolating currents. For a
reliable determination of the strong coupling constants of
$J^P={1\over2}^{\!-}$ octet baryons with pseudoscalar meson it is necessary
to find such regions of these parameters where coupling constant exhibits
good stability to the variation of them. Our analysis shows that the working
region of $M^2$ in the present work coincides practically with the one
determined for the magnetic moments of negative parity baryons in
\cite{Rvsf12}, as are given below,
\bea
&& 1.5\,GeV^2 \le M^2 \le 3.0\,GeV^2, \mbox{for $p^\ast$ and $n^\ast$}\,,\nnb\\
&& 1.8\,GeV^2 \le M^2 \le 3.5\,GeV^2, \mbox{for $\Lambda^\ast$, $\Sigma^\ast$
and $\Xi^\ast$}\,.\nnb
\eea

The continuum threshold $s_0$ is related to the first excited states. The
difference $\sqrt{s_0}-m_{ground}$ is the energy needed to transfer the
baryon to its first excited state. Analysis of all existing sum rules
approaches shows that this difference usually varies in the region from
$0.3\,GeV$ to $0.8\,GeV$, i.e., $m_{ground}+0.3\,GeV \le \sqrt{s_0} \le
m_{ground}+0.8\,GeV$, and in our analysis we chose the average value
$m_{ground}+0.5\,GeV$.

After deciding on the working regions of $M^2$ and $s_0$, our next and final
goal is to find the region of the auxiliary parameter $\beta$ where the
results are independent with respect to its variation.

As an example in Figs. 1 and 2 we present the dependence of
$g_{\Sigma^{0\ast} \Lambda^{0^\ast} \pi^0}$ coupling constant on $M^2$ at
four fixed values of $\beta$, and at $s_0=4.0\,GeV^2$ and $s_0=4.5\,GeV^2$,
respectively.
We see from these figures that the results are rather stable with respect to
the variation in $M^2$ and $\beta$.

In Figs. 3 and 4 we depict the dependence of the $g_{\Sigma^{0\ast}
\Lambda^{0^\ast} \pi^0}$ on $\cos\theta$ (where $\beta=\tan\theta)$ at three
fixed values of $M^2$, and at two fixed values of the continuum threshold
$s_0=4.0\,GeV^2$ and $4.5\,GeV^2$, respectively. We observe from these
figures that when $\cos\theta$ varies in the interval $-1.00\le \cos\theta
\le -0.85$ the coupling constant seems to be independent on the parameter
$\beta$ and we get $g_{\Sigma^{0\ast} \Lambda^{0^\ast} \pi^0}= 5.0 \pm 0.5$.

We perform similar analysis for the other coupling constants whose results
are all summarized in Table 1.


\begin{table}[h]

\renewcommand{\arraystretch}{1.3}
\addtolength{\arraycolsep}{-0.5pt}
\small
$$
\begin{array}{|l|c|c|}
\hline \hline
                            & \mbox{Negative parity} &
                              \mbox{Positive parity} \\
                            &\mbox{baryons}  & \mbox{baryons} \\  \hline
 \Lambda \to \Sigma^+ \pi^-   &    4   \pm 1   &   10   \pm 3    \\
 \Lambda \to \Xi^0 K^0        &    4   \pm 2   &    4.5 \pm 2.0  \\
 \Sigma^0 \to n K^0           &    3   \pm 1   &    4   \pm 3    \\
 \Sigma^0 \to \Lambda \pi^0   &    5.0 \pm 0.5 &   11   \pm 3    \\
 \Sigma^0 \to \Xi^0 K^0       &   10   \pm 2   &   13   \pm 2    \\
 \Sigma^- \to n K^-           &    3   \pm 1   &    5   \pm 3    \\
 \Sigma^+ \to \Lambda \pi^+   &    6   \pm 2   &   10.0 \pm 3.5  \\
 \Sigma^+ \to \Sigma^0 \pi^+  &    8   \pm 3   &    9   \pm 2    \\
 \Xi^0 \to \Lambda K^0        &    4.0 \pm 0.5 &    4.5 \pm 1.0  \\
 \Xi^0 \to \Sigma^0 K^0       &    9   \pm 1   &   12.5 \pm 3.0  \\
 \Xi^0 \to \Sigma^+ K^-       &   14   \pm 3   &   18   \pm 4    \\
 \Xi^0 \to \Xi^0 \pi^0        &    4   \pm 1   &   10   \pm 2    \\ 
\hline \hline  
\end{array}
$$
\caption{The strong coupling constants of negative and positive parity
octet baryons with pdeudoscalar mesons.}
\renewcommand{\arraystretch}{1}
\addtolength{\arraycolsep}{-1.0pt}
\end{table}


For completeness, in Table 2 we also present the modular values
of the strong coupling constants of the light pseudoscalar mesons
with positive parity octet baryons.

From the results given in Table 1 we deduce the following conclusions:
\begin{itemize}

\item In many cases the results obtained by using the general current differ
considerably from the predictions of the Ioffe current ($\beta=-1$). This difference can
be attributed to the fact that the coupling constant predicted by the Ioffe
current lies outside the stability region of $\beta$. As an example the
coupling constant is $g_{\Sigma^{0\ast} \Lambda^{0^\ast} \pi^0}=25$ for the
Ioffe current ($\beta=-1$), while it is $g_{\Sigma^{0\ast} \Lambda^{0^\ast}
\pi^0}=5.0\pm 0.5$ for the general current.

\item The values of the strong coupling constants for the negative and
positive parity baryons are close to each other in many cases. Considerable
difference occurs for the
$\Sigma^{+\ast} \to \Lambda^{0\ast} \pi^+$,
$\Sigma^{0\ast} \to \Lambda^{0\ast} \pi^0$,
$\Lambda^{0\ast} \to \Sigma^{+\ast} \pi^-$, and
$\Xi^{0\ast} \to \Xi^{0\ast} \pi^0$
channels.
 
\end{itemize}   

Finally it should be emphasized here that our prediction on the coupling
constant for the $N^\ast \to N^\ast \pi^0$ channel is approximately 50\%
larger than the one calculated in 3--point QCD sum rules approach
\cite{Rvsf28}.

In conclusion, the strong coupling constants of the light pseudoscalar
mesons with the negative parity octet baryons are calculated in framework of
the light cone QCD sum rules. We observe that all coupling constants
can be described only in terms of three invariant
functions, where two of then correspond to the well known $F$ and $D$
couplings in the $SU(3)_f$ symmetry, and the third one describes the $SU(3)_f$
violating effects. We present our predictions of the strong coupling
constants of $J^P={1\over 2}^{\,-}$ octet baryons and compare them with the
coupling constants of the corresponding positive parity baryons.    

\newpage
\section*{Appendix}
In this appendix we present the relations among the correlation functions in the
isospin symmetry limit. These relations hold for the positive and negative
parity baryons (see also \cite{Rvsf15}).\\
\begin{itemize}
\item Correlation functions involving the pions:
\bea
\Pi^{\Sigma^0 \to \Sigma^0 \pi} &=& \Pi^{\Lambda \to \Lambda \pi} = 0
\nnb \\
\sqrt2 \Pi_1(q,q,s) &=& \Pi^{\Sigma^+ \to \Sigma^+ \pi} = - \Pi^{\Sigma^- \to \Sigma^-
\pi} = - \Pi^{\Sigma^0 \to \Sigma^+ \pi} 
\nnb \\
&=& \Pi^{\Sigma^- \to \Sigma^0 \pi} = - \Pi^{\Sigma^+ \to \Sigma^0 \pi} = \Pi^{\Sigma^0 \to \Sigma^- \pi}
\nnb \\
\Pi^{\Xi^0 \to \Xi^0 \pi} &=& \frac1{\sqrt2} \Pi_2(s,s,q) = - \Pi^{\Xi^- \to \Xi^- \pi}= - \frac{1}{\sqrt2} \Pi^{\Xi^- \to \Xi^0 \pi}
= - \frac{1}{\sqrt2} \Pi^{\Xi^0 \to \Xi^- \pi}
\nnb \\
\Pi^{p \to p \pi} &=& - \Pi^{n \to n \pi} = \sqrt2 \Pi_1(q,q,q) - \frac1{\sqrt2} \Pi_2(q,q,q)
\nnb  \\
\Pi^{\Lambda \to \Sigma^+ \pi} &=& \Pi^{\Lambda \to \Sigma^- \pi} = -\frac1{\sqrt3} \left[ 2 \Pi_3(q,s,q) + \sqrt2 \Pi_1(q,q,s) \right]
\nnb \\
\Pi^{\Sigma^+ \to \Lambda \pi} &=& \Pi^{\Sigma^- \to \Lambda \pi} =- \frac1{\sqrt3} \left[ 2 \Pi_3(q,q,s) + \sqrt2 \Pi_1(q,q,s) \right]
\nnb \\
\Pi^{n \to p \pi} &=& \Pi^{p \to n \pi} = -\sqrt2\Pi_3(q,q,q)
\nnb \\
\Pi^{\Sigma^0  \to \Lambda \pi} + \Pi^{\Lambda \to \Sigma^0 \pi^0} &=& \frac{4}{\sqrt6} \left[ \Pi_1(q,s,q) - \Pi_2(s,q,q) \right] \nnb
\eea
\item {Correlation functions involving the kaons:}
\bea
\Pi^{n \to \Sigma^0 K} &=& - \Pi^{p \to \Sigma^0 K} = \Pi_3(s,q,q) + \sqrt2 \Pi_1(s,q,q)
\nnb \\ 
\Pi^{p \to \Lambda K} &=& \Pi^{n \to \Lambda K} = -\frac1{\sqrt3} \left[ \sqrt2 \Pi_1(s,q,q) - \Pi_3(s,q,q) \right]
\nnb \\
\Pi^{p \to \Sigma^+K} &=& \Pi^{n \to \Sigma^- K} = - \Pi_2(q,q,q)
\nnb \\
-\Pi^{\Sigma^0 \to \Xi^- K} &=& -\frac{1}{\sqrt2} \Pi^{\Sigma^+ \to \Xi^0 K} = \Pi^{\Sigma^0 \to \Xi^0 K} = -\frac{1}{\sqrt2} \Pi^{\Sigma^- \to
\Xi^- K} = \Pi_3(q,q,s)
\nnb \\
\Pi^{\Lambda \to \Xi^0 K} &=&  \Pi^{\Lambda \to \Xi^- K} = \frac1{\sqrt3} \left[ 2 \sqrt2 \Pi_1(q,s,q) + \Pi_3(q,q,s) \right]
\nnb \\
\Pi^{\Sigma^0 \to n K} &=& -\Pi^{\Sigma^0 \to p K} = \sqrt2 \Pi_1(s,q,q) + \Pi_3(s,q,q) 
\nnb \\
-\Pi^{\Lambda \to p K} &=& - \Pi^{\Lambda \to n K} = \frac1{\sqrt3}\left[ \sqrt2 \Pi_1(s,q,q) - \Pi_3(s,q,q) \right]
\nnb \\
\Pi^{\Sigma^- \to n K} &=& \Pi^{\Sigma^+ \to p K} = - \Pi_2(q,q,s)
\nnb \\
-\Pi^{\Xi^0 \to \Sigma^+ K} &=&- \Pi^{\Xi^- \to \Sigma^- K} = \sqrt2 \Pi_3(s,s,q)
\nnb \\ 
-\Pi^{\Xi^- \to \Sigma^0 K} &=& \Pi^{\Xi^0 \to \Sigma^0 K} = \Pi_3(q,s,q)
\nnb \\
\Pi^{\Xi^- \to \Lambda K} &=& \Pi^{\Xi^0 \to \Lambda K} = \frac1{\sqrt3} \left[ 2 \sqrt2 \Pi_1(q,s,q) + \Pi_3(q,q,s) \right]\nnb
\eea

\end{itemize}

\newpage

\section*{Figure captions}
{\bf Fig. (1)} The dependence of the strong coupling constant
$g_{\Sigma^{0\ast} \Lambda^{0\ast} \pi^0}$ on the Borel parameter $M^2$,
at the fixed value of the continuum threshold $s_0=4.0\,GeV^2$, and several
fixed values of  the auxliary parameter $\beta$.\\\\
{\bf Fig. (2)} The same as Fig. (1), but at the fixed value of the
continuum threshold $s_0=4.5\,GeV^2$.\\\\
{\bf Fig. (3)} The dependence of the strong coupling constant
$g_{\Sigma^{0\ast} \Lambda^{0\ast} \pi^0}$ on $\cos\theta$, at the fixed value of the
continuum threshold $s_0=4.0\,GeV^2$, at several fixed values of $M^2$.\\\\
{\bf Fig. (4)} The same as Fig. (3), but at the fixed value of the
continuum threshold $s_0=4.5\,GeV^2$.

\newpage

\begin{figure}
\vskip 3. cm
    \includegraphics{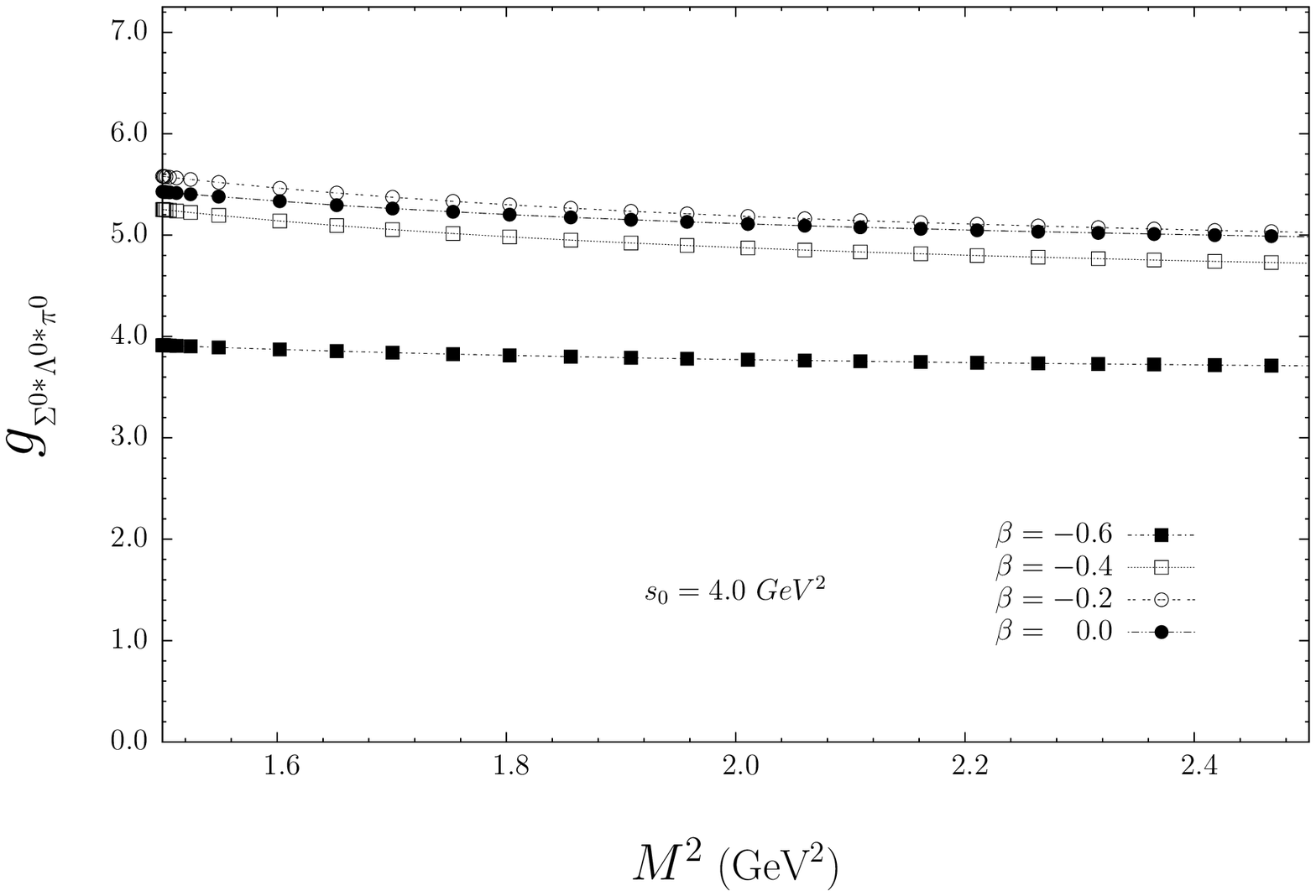}
\vskip 7.0cm
\caption{}
\end{figure}

\begin{figure}
\vskip 3. cm
    \includegraphics{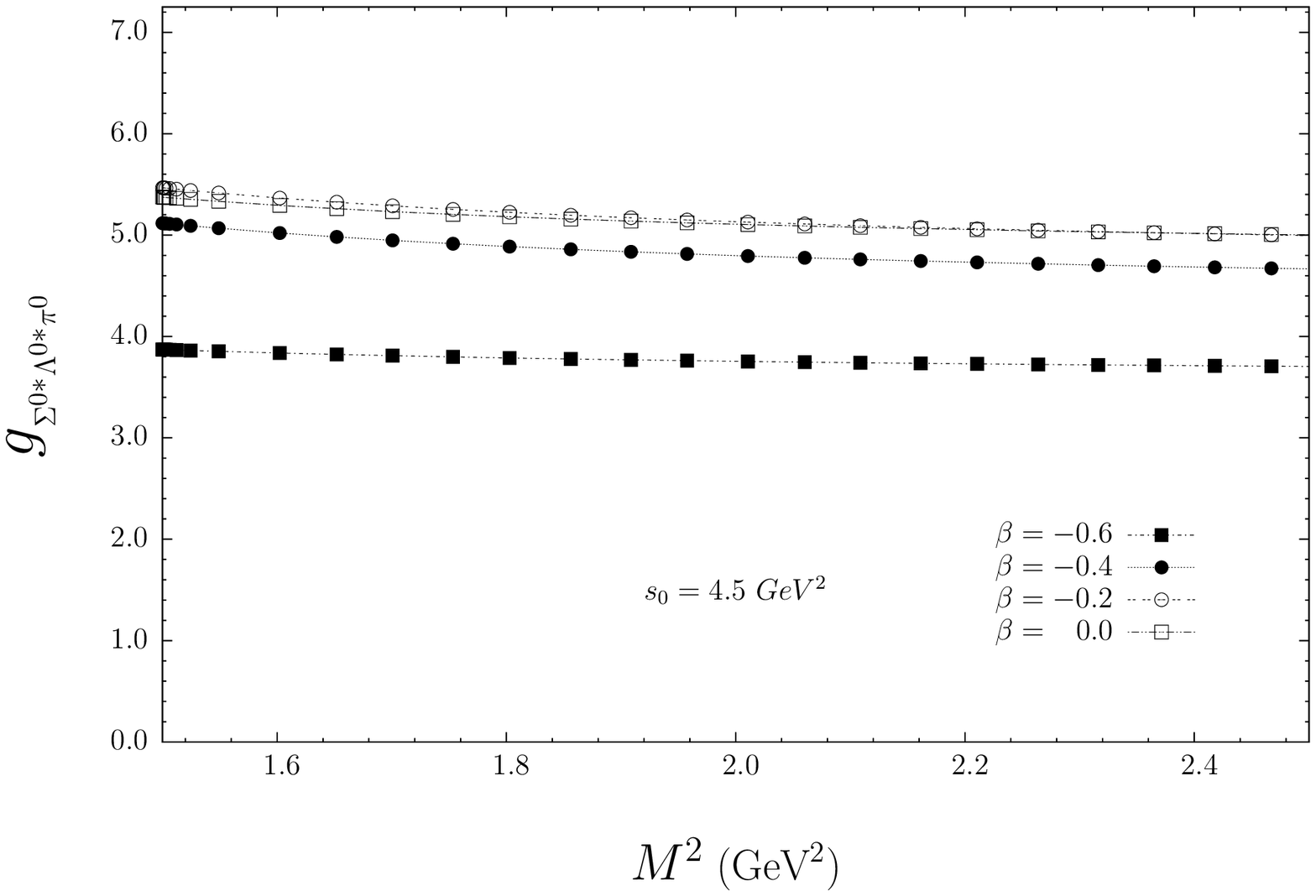}
\vskip 7.0cm
\caption{}
\end{figure}

\begin{figure}
\vskip 3. cm
    \includegraphics{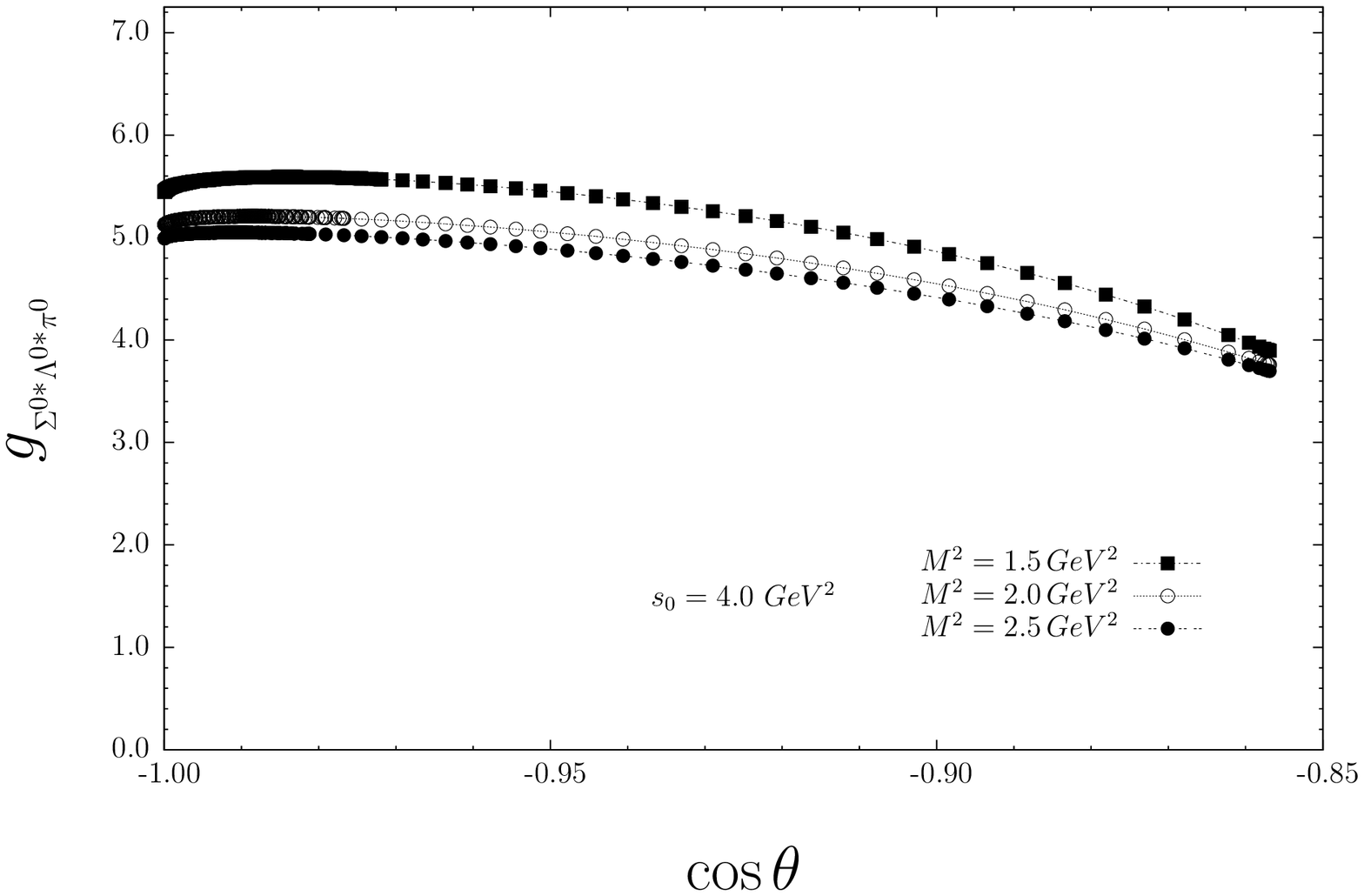}
\vskip 7.0cm
\caption{}
\end{figure}

\begin{figure}
\vskip 3. cm
    \includegraphics{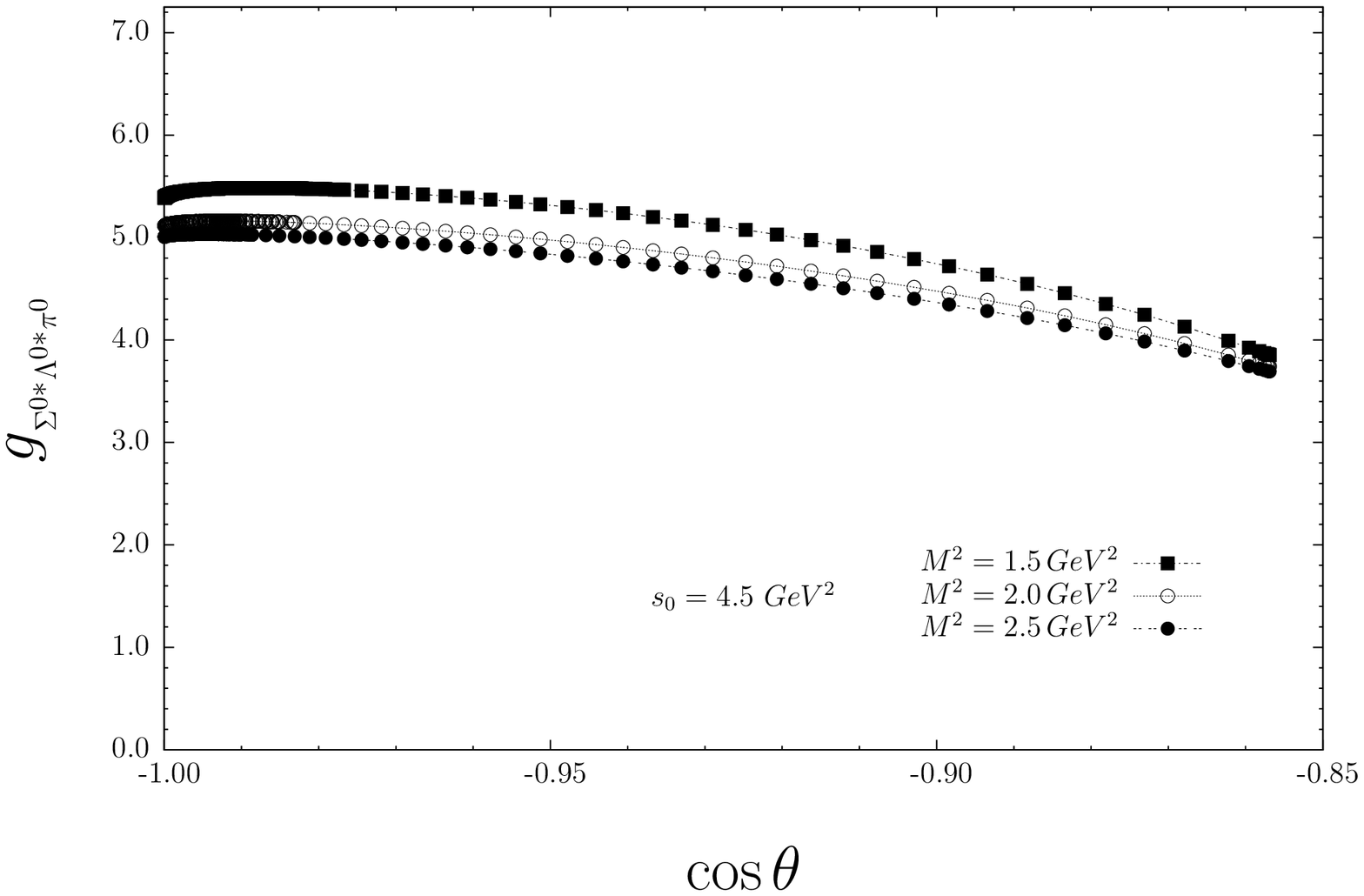}
\vskip 7.0cm
\caption{}
\end{figure}

\end{document}